# Design and performance of an ultra-high vacuum spin-polarized scanning tunneling microscope operating at 30 mK and in a vector magnetic field

Henning von Allwörden[1], Andreas Eich[1], Elze J. Knol[1], Jan Hermenau[2], Andreas Sonntag[2], Jan W. Gerritsen[1], Daniel Wegner[1], and Alexander A. Khajetoorians[1]

[1] *Scanning Probe Microscopy Department, Institute for Molecules and Materials, Radboud University, Nijmegen, The Netherlands*

[2] *Department of Physics, Hamburg University, Hamburg, Germany*

We describe the design and performance of a scanning tunneling microscope (STM) which operates at a base temperature of 30 mK in a vector magnetic field. The cryogenics is based on an ultra-high vacuum (UHV) top-loading wet dilution refrigerator that contains a vector magnet allowing for fields up to 9 T perpendicular and 4 T parallel to the sample. The STM is placed in a multi-chamber UHV system, which allows *in-situ* preparation and exchange of samples and tips. The entire system rests on a 150-ton concrete block suspended by pneumatic isolators, which is housed in an acoustically isolated and electromagnetically shielded laboratory optimized for extremely low noise scanning probe measurements. We demonstrate the overall performance by illustrating atomic resolution and quasiparticle interference imaging and detail the vibrational noise of both the laboratory and microscope. We also determine the electron temperature via measurement of the superconducting gap of Re(0001) and illustrate magnetic field-dependent measurements of the spin excitations of individual Fe atoms on Pt(111). Finally, we demonstrate spin resolution by imaging the magnetic structure of the Fe double layer on W(110).

**Introduction**

The development of scanning tunneling microscopy (STM) in ultra-high vacuum (UHV) and magnetic fields at sub-Kelvin temperatures ($T \leq 1$ K)[1-3] has enabled new classes of studies toward understanding the magnetism of single and coupled magnetic atoms on non-magnetic surfaces. For example, spin-based inelastic tunneling spectroscopy [4] has enabled the characterization of magnetic anisotropy of individual magnetic atoms [5-7] and molecules [8, 9] on a variety of surfaces. Moreover, it has been utilized to characterize various types of exchange couplings in bottom-up constructed chains [10-13]. Complementary to inelastic tunneling, spin-polarized scanning tunneling spectroscopy (SP-STS) has been applied to magnetic atoms in magnetic fields on non-magnetic surfaces to probe the magnetization of individual and coupled magnetic adatoms [6, 14-17]. More recently, these methods have been combined with microwave sources in pump-probe schemes to measure the various lifetimes of magnetic adatoms on thin film insulators [18, 19], as well as perform electron spin resonance on an individual atom [20, 21].

While sub-Kelvin STM in magnetic fields based on $^3$He cryogenics [1] has vastly opened new fields, the lowest possible temperature limits its application to look at novel phases of matter, to probe with ultrahigh-energy resolution [22-25], or to conduct experiments that have stringent requirements on cooling power or hold time. Over the last years, there has been a strong drive toward the development of UHV-compatible dilution refrigeration STM systems, in order to push the experimentally accessible temperatures into the tens of milli-Kelvin range [3, 26, 27]. However, implementation of these systems is extremely challenging due in part to: (1) the continuous flow of cryogens based on a gas handling system (GHS) necessitating good decoupling from the microscope environment; (2) the complexity of the cryogenics combined with the need for UHV-type seals as well as transfer space for sample and probe exchange; (3) reaching electron temperatures near the base temperature of the dilution refrigerator. Concerning the electron temperature [28], it has yet to be convincingly demonstrated that electron temperatures in the range of the base temperature can be achieved with an STM setup.

We describe the design and performance of a scanning tunneling microscope (STM) that operates at a base temperature of 30 mK and in vector magnetic fields. The cryogenics are based on a commercially available compact ultra-high vacuum top-loading dilution refrigerator that houses a 2D vector magnet allowing for fields up to 9 T perpendicular and 4 T parallel to the sample. The STM is designed based on a Pan-style motor and rigidly connected to the mixing chamber. The microscope is coupled to a multi-chamber UHV system, which allows for *in-situ* preparation and exchange of samples as well as tips. Therefore, the UHV system is equipped for molecular beam epitaxy (MBE) and chemical vapor deposition (CVD), usable at variable sample temperatures ranging from 10 to 2000 K. Moreover, both samples and tips can be pre-cooled via a vertical manipulator containing a continuous-flow He cryostat, and then directly transferred into the microscope in order to limit temperature-induced diffusion of deposited materials. The system is housed in an acoustically isolated and electromagnetically shielded laboratory, where the lab floor is made from a 150-ton concrete block suspended by pneumatic isolators, all being optimized for extremely low-noise scanning probe measurements. We detail the vibrational performance of the laboratory, utilizing highly sensitive accelerometers. As a demonstration of the microscope's performance, we illustrate the atomic resolution and quasiparticle interference of various surfaces and analyze the vibrational and electronic noise. We also determine the effective electron temperature via analysis of superconducting gap spectra, and present inelastic tunneling spectroscopy of individual Fe atoms on Pt(111) in magnetic field [29]. Finally, we demonstrate spin resolution by imaging the magnetic structure of the Fe double layer on W(110), which is known to have out-of-plane magnetized domains, with in-plane magnetized domain walls [30].

## I. Still Lab

The "Still" laboratory was specifically designed and built to optimize the vibrational performance and to minimize the influence of electromagnetic radiation for high-resolution, dilution-refrigerator based microscopes [26, 27, 31]. As the STM junction is exponentially sensitive to atomic-scale variations of the tip-sample separation, the laboratory was constructed utilizing the concept discussed in Ref. [32], trying to isolate the various eigenfrequencies of the laboratory and facility in order to minimize their overlap. Figure 1 shows a 3D-rendered overview of the laboratory layout, including a model of the UHV chamber system with attached cryostat (see section II).

*a. Lab design*

The laboratory is divided into three separate rooms, namely the facility space, the pump room for the gas handling system, and the data acquisition area. The laboratory is located three floors underground and its deepest point reaches about 15 m deep. The facility space is comprised of a 150-ton solid, non-magnetic concrete block, reinforced by fiber glass. The block has a surface area of 5.6 x 4.8 m² and is 4 m high. Additionally, there is a pit in the center of the block with a depth of 2.4 m and area of 1.2 x 2.2 m². The block rests on six pneumatic passive dampers (Bilz BiAir® 5-HE-MAX) with a vertical/horizontal resonance frequency of 1.2/2.5 Hz, respectively. The pneumatic dampers stand on concrete pillars that are rigidly connected to the side walls, and both are rigidly connected to the isolated foundation, which rests on sand and is not connected to the surrounding building. Thus, up to the lab's floor level (i.e. the top of the block), the pillars, walls and foundation of the facility space constitute a rigid object. The walls above lab-floor level and the ceiling are connected to the lower base via a 2-cm thick Silomer® layer to reduce coupling of vibrations from the upper half of the room to the foundation. Hence, the entire room is isolated from the building. The walls contain acoustic paneling designed for damping of airborne vibrational noise. The whole volume of the facility space is completely enclosed by steel plates, creating a Faraday cage for electromagnetic shielding.

The usage of a dilution refrigerator requires a gas handling system, which contains a roots pump to circulate the helium mixture while the experimental setup is at base temperature (see section III). In order to minimize vibrations induced by the gas handling system, a pump room with additional acoustic paneling was constructed next to the facility space. The necessary pipe connections are made via ports through the intermediate double wall, but otherwise the two spaces are completely decoupled from each other. The pipe joints are decoupled via a gimbal system (McAllister Technical Services (USA)) from the facility wall and the floating block [26, 27].

All of the necessary electronics needed for use of the UHV-system are placed inside the facility space. All data acquisition hardware is accessed through electromagnetically shielded ports, similar to the paneling for the gas handling system, and signals are acquired via network cables enabling measurement in an external office space. All the signal lines are low-pass filtered by different combinations of low-pass and EMI filters, as discussed in section V.c, to maximize the energy resolution.

*b. Lab Performance*

To test the vibrational noise level of the facility space, measurements with piezoelectric seismic accelerometers (Wilcoxon Research (USA), accelerometers 731A; sensitivity 10 V/$g$; transverse sensitivity max. 1% of axial) were taken. The charge signal, produced by an acceleration acting on the devices, is amplified, filtered and converted into a voltage in corresponding units. The continuous, time-dependent data is sent to a computer, which processes it with a home-built, LabVIEW-based software. It integrates the signal to get a time-dependent velocity, and by applying a discrete Fourier transformation extracts the frequency-dependent spectral density. Figure 2 illustrates the measurements, where the three shown data sets are related to different measurement positions in the facility space: (i) the underlying foundation (red line), as well as the 150-ton concrete block (ii) non-floating (dark blue) and (iii) floating (bright blue), respectively. The intrinsic damping of the block due to its weight and stiffness is visible for frequencies above 20 Hz, lowering the spectral density close to 100 Hz nearly into the noise floor of the detection equipment. The effect of floating the block using the air springs vastly improves the damping of vibrational noise in the range of a few Hertz, reducing it over a large range nearly down to the noise floor. Only at low frequencies below 5 Hz the resonance of the dampers itself and vibrational eigenmodes of the concrete block are visible.

Two more frequency regions can be related to the floating block. First, the peak around 15 Hz is due to a standing-wave mode of the circulating air in the room. We can controllably dampen this peak by decelerating the airflow from the foundation to the ceiling (see Figure 1). We use noise protection foam panels at half room height (corresponding to the top of the 150-ton block), where the waves have their antinodes. The optimized damping situation is illustrated in Figure 2. Secondly, peaks slightly below and at 50 Hz are measured. While the power frequency of 50 Hz cannot be completely avoided, the peaks around 48 Hz are generated by ground-borne vibrational noise produced by powerful water pumps of the nearby high field magnet laboratory (HFML).

## II.     UHV System

The UHV system (Figure 3) is comprised of three separate main chambers and a fast load-lock chamber. Tips and samples can be introduced into and prepared within the UHV environment, transferred between different chambers and also into or out of the microscope *in-situ.* The chambers are mounted on a customized aluminum frame. Bake-out of the UHV section is achieved with a customized tent (Hemi Heating (Sweden)), which enables a constant temperature, regulated up to 150°C. The base pressure of the UHV chambers after bake-out is typically below $2 \cdot 10^{-10}$ mbar, measured with Bayard-Alpert gauges (Vacom (Germany), BARION®).

The two satellite UHV chambers are dedicated to sample and tip preparation. One UHV chamber is used for different cleaning procedures such as direct ion bombardment (sputtering) or heating to 2400 K by electron-beam bombardment from the sample backside. Moreover, this chamber is equipped with leak valves that enable precise dosage of gases and thus CVD growth of various materials, such as graphene or single-layer transition metal dichalcogenides [33-35]. Since these processes require higher pressures, a second UHV chamber is dedicated to sample annealing at base pressure as well as MBE growth. MBE can be performed via e-beam evaporators or Knudsen cells on both tips and samples for temperatures ranging from 80 - 1000 K.

The central UHV chamber is used for sample and tip storage with 20 places, their transfer into and out of the microscope, as well as deposition of materials near $^4$He temperatures. A vertical manipulator is used for the transfer, which contains a He-flow cryostat directly attached to the transfer stage. This setup enables cool down of tips and samples to 4.3 K, either to precool them before a transfer into the cold environment of the cryostat or for cold deposition of e.g. single atoms via an aligned evaporator (see section V.d). The vertical manipulator is custom-designed, based on the KONTI cryostat design from Cryovac GmbH & Co. KG (Germany). The transfer between the chambers is done by magnetically coupled linear translators. Samples and tips are handled in each chamber with wobble sticks to give full flexibility and access to all the described preparation stages. Each UHV chamber is pumped by its own ion-getter pump for quiet operation during measurements, while turbo molecular pumps are attached on the two satellite chambers.

The cryostat includes the dilution refrigeration insert (section III) and is mounted underneath the central UHV chamber. It extends into a pit, as displayed in Figure 1, and can be lowered utilizing a pulley-winch system without the need to detach the insert from the UHV system. The insert is connected to the central UHV chamber via a gate valve, enabling easy disconnection of the cryogenic system without disturbing the UHV chambers. This enables quick access to the STM when the system is at room temperature.

### III.    Cryogenics

Our system utilizes a commercial wet dilution refrigeration unit (Janis Research Company (USA), JDR-50) in order to access temperatures in the range of tens of milli-kelvin. It is fully UHV compatible and top loading (Figure 4 (a)). The insert is placed inside a $^4$He bath cryostat equipped with a vector magnet (up to 9 T vertically and up to 4 T along one horizontal axis; vector fields at 45° can be up to 3 T). The dewar is super-insulated without the need for a liquid nitrogen shield, hence preventing commonly known noise problems from $LN_2$ boil-off. Its helium belly has a capacity of ca. 80 liters, yielding a hold time of more than 4 days. The UHV space is separated from the $^4$He bath by a so-called inner vacuum can (IVC), which is sealed via a conflat flange, immersed into the liquid $^4$He.

The dilution unit and attached microscope are all located inside the IVC (Figure 4 (b)). Precooling is achieved via a mechanical heat switch which creates thermal contact to the $^4$He bath via cold finger. The mechanical heat switch is additionally connected to two rotatable shutters, which enable thermal contact and shielding of various parts of the dilution unit from black-body radiation within the transfer space (Ø 38 mm). Both shutters have electrical contacts, enabling indication whether the shutters are fully open or closed. Furthermore, all parts underneath the still are fully surrounded by a 1 K radiation shield which is directly connected to the still.

The insert is equipped with a variety of cabling leading from the feedthrough to the mixing chamber: (1) NbTi twisted pairs with a CuNi clad and a stainless steel shield, used for temperature sensors, cartridge heaters (only from still to mixing chamber) and high voltages for piezos; (2) spare stainless steel coaxial cables; (3) NbTi single strands with a CuNi clad and a stainless steel shield for shutter indicator and piezoelectric stepper motor (walker); (4) stainless steel and NbTi semi-rigid cables for bias voltage and tunnel current; and (5) phosphor bronze twisted pairs non-shielded for the cartridge heaters from feedthrough to the still. From the mixing chamber to the STM, homemade twisted-pair cabling from polyimide-coated silver wire is used, as well as copper semi-rigid cables for the voltage and current signal lines and shielded superconducting twisted pairs for the temperature sensor.

The cooling power, which was measured without the STM unit attached, is 115 µW (95 µW) at a temperature of 100 mK (see Figure 5) with (without) continuously running compressor needed for the Joule-Thomson cooling stage (see below). These measurements can be done by introduction of heat on the still and mixing chamber with two separate cartridge heaters, which also allow experiments at elevated, stable temperatures of up to 600 mK. Temperatures at the IVC flange and the still stage are measured using calibrated Cernox sensors (Lake Shore Cryotronics (USA)), while calibrated ruthenium oxide ($RuO_2$) sensors (based on Lake Shore Rox™ sensors, calibrated by Janis Research Company (USA)) are used for base temperature measurements at the mixing-chamber stage and next to the STM head. Starting from 7 K, the unit with the microscope attached needs approximately one hour to reach $T =$ 40 mK at the mixing chamber.

The working principle of a dilution refrigerator is based on the different quantum-mechanical behavior of the two isotopes $^3$He and $^4$He used together in a mixture. If cooled below 600 mK, they separate into two phases, namely a concentrated and at low temperatures purified $^3$He phase and a diluted $^3$He/$^4$He phase. By extraction of mainly $^3$He, a crossing of He atoms over the phase boundary is induced, resulting in a cooling due to the different enthalpies [36]. To get below 600mK, the helium mixture is precooled via the $^4$He bath followed by a Joule-Thomson stage. This enables liquification of the mixture at a temperature of roughly 1.2 K. Subsequently, it passes a coiled and a silver-powder heat exchanger by which additional cooling from the outgoing colder mixture is present reducing the overall temperature of the mixture below 100 mK. In the final step, the mixture enters the mixing chamber, where the phase separation is located. The final cooling effect is promoted by the crossing of $^3$He isotopes from the rich to the diluted phase. To make this a continuous process, mainly $^3$He is extracted from the diluted phase by slightly heating the still, located above the mixing chamber. This distillation process happens due to the higher vapor pressure of $^3$He compared to $^4$He at a temperature of roughly 900 mK. Finally, the exhaust is pumped out with an external roots pump, and the warm mixture is fed back into the dilution unit via a closed cycle. In order to clean the mixture, it flows through a cold trap filled with activated carbon and inserted in a liquid nitrogen bath. The roots pump is located in the GHS, which also contains a dump for the mixture, needed if the system is not at base temperature, as well as several valves for control of mixture flow and a compressor for the JT cooling stage.

## IV. STM

The microscope design is based on the Pan concept with a vertically driven (*z*-axis) piezo-electrical stepper motor (walker) and can be later upgraded with an *x-y*-translational stage. The Pan concept was chosen due to its robust design and relatively high resonance frequencies [37, 38], which allows us to implement a head design without additional *in-situ* spring suspension. The microscope body is home-built, and special care was taken to choose materials that are non-magnetic and UHV-compatible at milli-Kelvin temperatures. A silver-tungsten sintered alloy was chosen for the microscope body, due to its relative high stiffness and high thermal conductivity. Typically, PEEK, alumina, and Shapal™ ceramics were used for insulating components, and all fasteners were made from phosphor bronze or PEEK. The piezoelectric tube scanner (EBL products (USA), PZT-5) has a scan range at base temperature of roughly $x,y = \pm 1$ µm and $z = \pm 225$ nm by using a voltage of $U = \pm 220$ V.

The microscope, shown as a cut-view CAD model in Figure 6, is very compact with an outer diameter of 39 mm and a total height of 58 mm only. The microscope is fastened to a silver plate, which is directly connected to the mixing chamber via four segmented silver posts. The plate exhibits a home-made plugging mechanism for the wiring, allowing easy and reliable disconnection and reconnection of the microscope if needed. All described silver parts are gold-coated to prevent oxidization. The calibrated RuO$_2$ temperature sensor is attached to the silver plate right next to the STM head.

As there is no optical access to the microscope in operation, the tip and sample transfer is done *in situ* utilizing a bayonet-type principle. Therefore, the aforementioned vertical motor-driven manipulator (see section II) from the top is used, which hosts different transfer shuttles for either tip or sample. In order to remove the tip, the sample must be removed first. With tip and sample placed in the microscope, the dimensions of the different parts ensure that there are no open gaps by which radiation could couple directly into the tip-sample junction. For data acquisition, a Nanonis RC5/SC5 setup (SPECS Zurich (Switzerland)) with piezo motor driver PMD4 and high-voltage power supply HV4 in combination with software version v4.5 is used. For the current measurement we use a commercially available pre-amplifier Femto DLCPA-200 (FEMTO Messtechnik (Germany)).

### V. Performance

In order to ascertain the overall effect of the laboratory damping, we measured the electronic and vibrational performance of the tip-sample junction in tunneling. Below, we detail the overall performance using different measurement methods.

*a. I/Z noise*

To characterize the vibrational noise at the tunneling junction, the spectral density of the tunnel current was measured with the acquisition electronics (Figure 7), taken at a mixing chamber temperature of $T_{MC}$ = 43 mK and for frequencies up to 1 kHz. For the measurement, a PtIr tip was placed over a thick Pb film grown on a W(110) crystal. The W(110) was cleaned by several cycles of annealing in oxygen atmosphere at 1100°C and subsequent flashing to 2300°C for several seconds. Out of tunneling range, i.e. retracted tip, the intrinsic electronic noise level is visible (green curve in Figure 7 (a)). After approaching the tip, there is an increase of the noise level by a factor of 10 (blue), except for low frequencies which are stabilized by the feedback circuit. The overall spectral density stays around 60 fA Hz$^{-½}$ which is very low and comparable or even slightly better compared to similar state-of-the-art setups [26, 27, 39, 40]. This is also true for the measured spectral density of the *z*-height (Figure 7 (b)), which stays well below 20 fm Hz$^{-½}$ and reaches 5 fm Hz$^{-½}$ for frequencies above 400 Hz. From comparison between the damping characteristics of the floating floor and that of the STM, the noise level on the concrete block (up to 100 Hz) is two magnitudes better than the laboratory foundation (Figure 2 and section I.b), while the STM is improved only one order of magnitude. We also checked the current noise in tunneling regime under open-feedback conditions (setpoint values prior to opening: $I_t$ = 500 pA, $V_S$ = 10 mV, $T_{MC}$ = 43 mK). This corresponds to the input noise when acquiring scanning tunneling spectroscopy (STS) spectra or maps. Additionally, the spectral density is given in Figure 7 (a) (red), which is from frequencies of about 400 Hz on only a factor of two above the total noise level with values around 30 fA Hz$^{-½}$, allowing detection of very small signal changes.

### b. Atomic resolution

In order to further test the imaging performance of the microscope, we demonstrate high lateral resolution on a Cu(111) surface. The surface was prepared by repeated argon ion bombardment and subsequent annealing at 560°C. After the transfer into the microscope and cool-down to base temperature ($T_{MC}$ = 29 mK), atomic resolution of the surface was obtained with a PtIr tip (**Error! Reference source not found.** (a)). Atomic-scale images at low bias ($V_s$ = 12 mV) nicely reveal the three-fold symmetric (1x1) surface structure of the crystal. Additionally, the quasi-particle interference (QPI) of the surface-state electrons at the Fermi energy can be seen in this figure, identified as standing-wave pattern [41][42, 43]. With the present setup, images of this quality are typically obtained without the need for long-time averaged scans, or the need for heavy post-image filtering.

### c. Superconducting gap

In order to ascertain the overall electron temperature of the tunnel junction, *I-V* spectroscopy in open feedback was performed using a lock-in technique. We utilized Re(0001), which has an energy gap $\Delta$ = 255 µeV and $T_C$ = 1.6 K [44]. While multiple systems have demonstrated base temperatures in the 10 mK range [3, 27, 40, 45-47], there is still an open question if the electron temperature can reach base temperature, or if there are other limitations. As the junction is sensitive to RF noise, we utilized low-pass filtering on all lines. In addition, we also included additional state-of-the-art powder filters on the bias line [48]. In order to minimize the effect of 50 Hz noise, which is the power frequency, we utilized a star-ground layout, where the pre-amplifier for the tunnel current is the central star point. The measurements were done with a lock-in amplifier using an AC modulation of the applied bias voltage of $V_{mod}$ = 4 µV at a frequency of $f_{mod}$ = 821 Hz while the setup was at base temperature $T_{MC}$ = 47.5 mK. The result is shown in Figure 9 together with a fit using the Dynes equation for a BCS superconductor [49, 50]. Considering a lifetime broadening $\Gamma$ = 0.005 mV, we acquire an effective electron temperature of $T_{eff}$ = 195 mK. We note that different fitting functions are utilized in literature, which gives slightly different effective temperatures, e.g. via the Maki equations [3, 51].

### d. Magnetic field-dependent spin-excitation measurements

In order to demonstrate measurements in magnetic field, we measured the spin excitations of individual Fe atoms on Pt(111) [29, 52]. It was shown that single Fe atoms adsorb on hollow sites (fcc/hcp) of Pt(111) when cold-deposited onto the surface. Fe was deposited on a Pt(111) surface with an e-beam evaporator, while the sample was cooled to $T$ = 4.3 K via our vertical manipulator. The sample was continuously cooled and subsequently inserted into the microscope, revealing the presence of single Fe atoms, both clean and hydrogenated as discussed in ref. [52], with an apparent height of roughly 130 pm and 200 pm, for the clean and single hydrogenated case respectively (Figure 10 (a)). Tunneling spectroscopy (Figure 10 (b)) performed in open-feedback conditions with a Cr bulk tip ($V_{stab}$ = 5, 10 mV, $I_{stab}$ = 44, 500 pA, $f_{mod}$ = 781 Hz, $V_{mod}$ = 60 µV$_{rms}$) reveal inelastic tunneling features at $V$ = 0.6 mV (0.1 mV), which are related to the previously reported spin excitations of fcc (hcp) Fe atoms, respectively. In order to remove features due to the tip, we normalize the raw spectra by dividing the Fe spectra with spectra taken on the substrate. Individual Fe atoms exhibit a hollow-site dependent magnetic anisotropy, where fcc atoms exhibit an out-of-plane easy axis (high spin), and hcp atoms exhibit an easy in-plane (low spin). This site dependence explains the difference in spin excitations for the two binding sites. Moreover, we confirmed this dependence by performing atomic manipulation of a single atom, and observed the switching of the excitation behavior as illustrated.

In order to prove the magnetic nature of the inelastic excitations, we applied an out-of-plane magnetic field ($B_z$ = 8 T), to observe the Zeeman shift of the excitation. As seen in ref. [29], we observe a change of energy of the excitation to 1.2 mV and 0.4 mV for the fcc and hcp site, respectively. The larger shift of the fcc atom with respect to the hcp atom is due to the difference in ground-state orientations. Additionally, we observe a broadening of the linewidth of the excitation, similar to what was previously observed.

### e. Spin-polarized STM of 2 ML Fe/W(110)

To demonstrate magnetic resolution, we characterized the Fe double layer grown on W(110) with a bulk Cr tip. The Cr tip was prepared from a wire with 0.5 mm diameter utilizing the etching procedure described in Ref. [53] and cleaned *in-situ* by field emission until spin contrast was achieved. The double layer of Fe was prepared on W(110) applying the procedures described in Refs. [30, 54]. Annealing the sample at sufficient temperature leads to step-flow growth of Fe on the step edges of W(110), as seen in Figure 11 (a). A signature of the double layer of Fe is the appearance of dislocation lines, which are oriented perpendicular to the domain walls of the double layer. The domain walls can be imaged at $V_S$ = 50 mV without magnetic sensitivity (Figure 11 (b)), allowing for easy identification of the various magnetized domains [55]. We utilize a Cr tip with a canted out-of-plane spin polarization, which allows imaging of both the domain walls and the domains at particular voltages [30]. Figure 11 (c) illustrates a two-tone contrast between alternating domain walls ($V_S$ = 15 mV), indicative of the in-plane spin polarization of the rotating domain walls [54]. The domain walls separate domains that are out-of-plane magnetized, which can be imaged with a tip with a net out-of-plane spin polarization. Figure 11 (d) illustrates the typical alternating contrast between subsequent domains ($V_S$ = -10 mV). All images were taken with an identical unmodified tip.

### VI.    Conclusion

In conclusion, we demonstrate the functionality of a scanning tunneling microscope operating at 30 mK and enclosed in a UHV system, in an ultra-quiet laboratory. From accelerometer measurements, we illustrate that the vibrational noise of the laboratory is similar to many recently reported labs. Furthermore, we demonstrate that the vibrational performance of the microscope in tunneling is suitable for atomic manipulation and high-resolution spectroscopic experiments. To demonstrate the energy resolution, we measured the superconducting gap from tunneling spectroscopy of Re(0001). We find that the effective energy resolution is approximately 200 mK, even though we introduce cold filtering of the bias line, and utilize superconducting cabling for the current line which has an inherent cutoff frequency of 16.8 MHz due to the skin effect [56]. One point of improvement would be to filter all lines at low temperature, and implement a cold preamplifier for the tunneling current measurement. As a demonstration of high-resolution measurements in magnetic field, we illustrate spin excitations of individual Fe atoms on the Pt(111) surface, which shows comparable results to previous reports [29, 52]. Finally, we illustrate spin contrast of the domains and domain walls within the double layer of Fe on W(110), showing the capability of spin-polarized STM at the system base temperature of 30 mK.

## Acknowledgements

We would like to acknowledge discussions with Steffen Wirth, Doug Bonn, Joseph Stroscio, Christian Ast, Ali Yazdani, and Vladimir Shvarts. We would also like to thankfully acknowledge scientific discussions with Nadine Hauptmann. We thank the TechnoCenter of the Faculty of Science at Radboud University for their technical support. We also appreciate discussions and support by Ulf Motz (Bilz Vibration Technology AG, Germany). We would like to acknowledge financial support from the Emmy Noether Program (KH324/1-1) via the Deutsche Forschungsgemeinschaft, the Foundation of Fundamental Research on Matter (FOM), which is part of the Netherlands Organization for Scientific Research (NWO), and the VIDI project: 'Manipulating the interplay between superconductivity and chiral magnetism at the single atom level' with project number 680-47-534 which is financed by NWO.

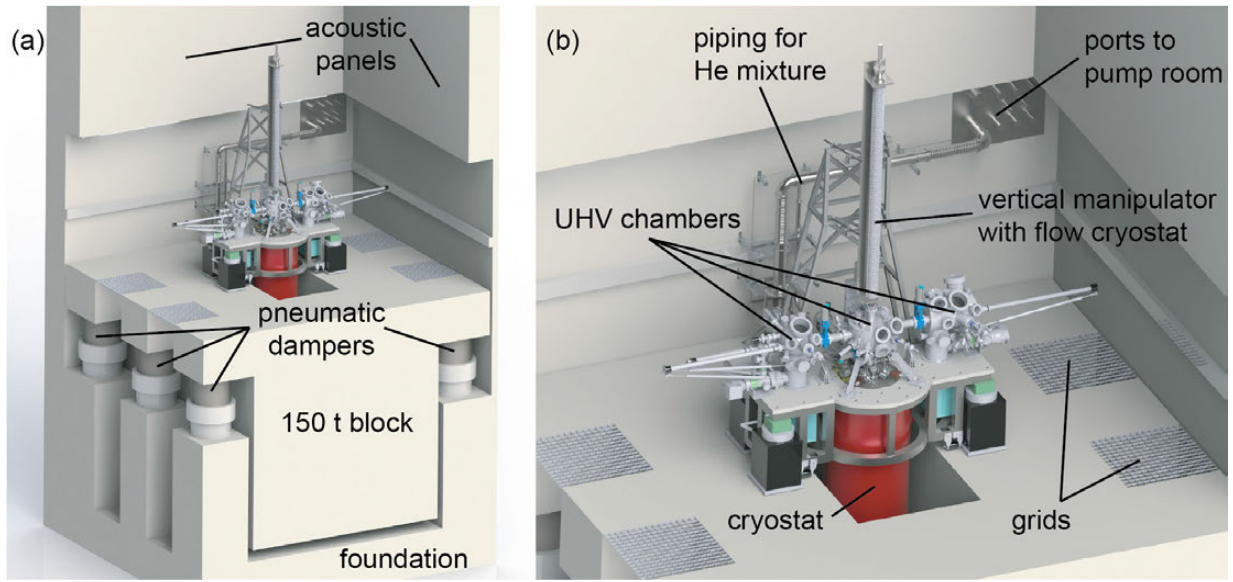

Figure 1. Layout of the STILL laboratory. (a) The 5.6 x 4.8 m² floor of the lab itself consists of a 150 ton concrete block resting on six pneumatic passive dampers. The whole space is surrounded by a Faraday cage made from steel plates and is not connected to the remaining building. (b) The UHV-system with the attached top-loading cryostat (red) is placed inside the lab as pictured. In the back the piping and damping system utilized for the closed He-mixture cycle is visible, which leads through the lab wall and connects to the gas handling system in a separate room (not illustrated).

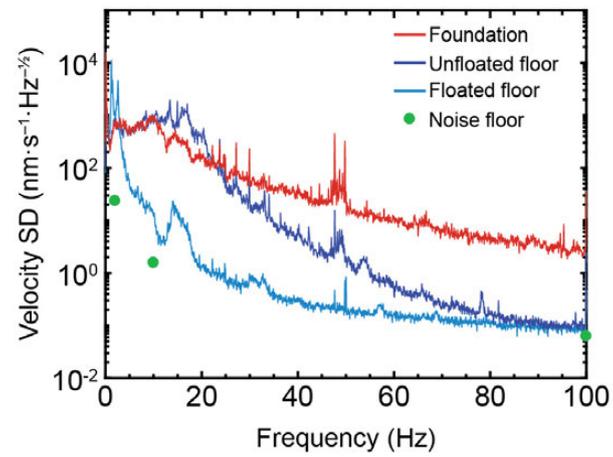

Figure 2. Vibrational noise measurements of the lab, measured on the foundation (red), on the unfloated (dark blue) and floated concrete block (bright blue), respectively. Shown is the velocity spectral density. In addition, the noise floor (green dots) of the used accelerometer in combination with the spectrum analyzer is shown.

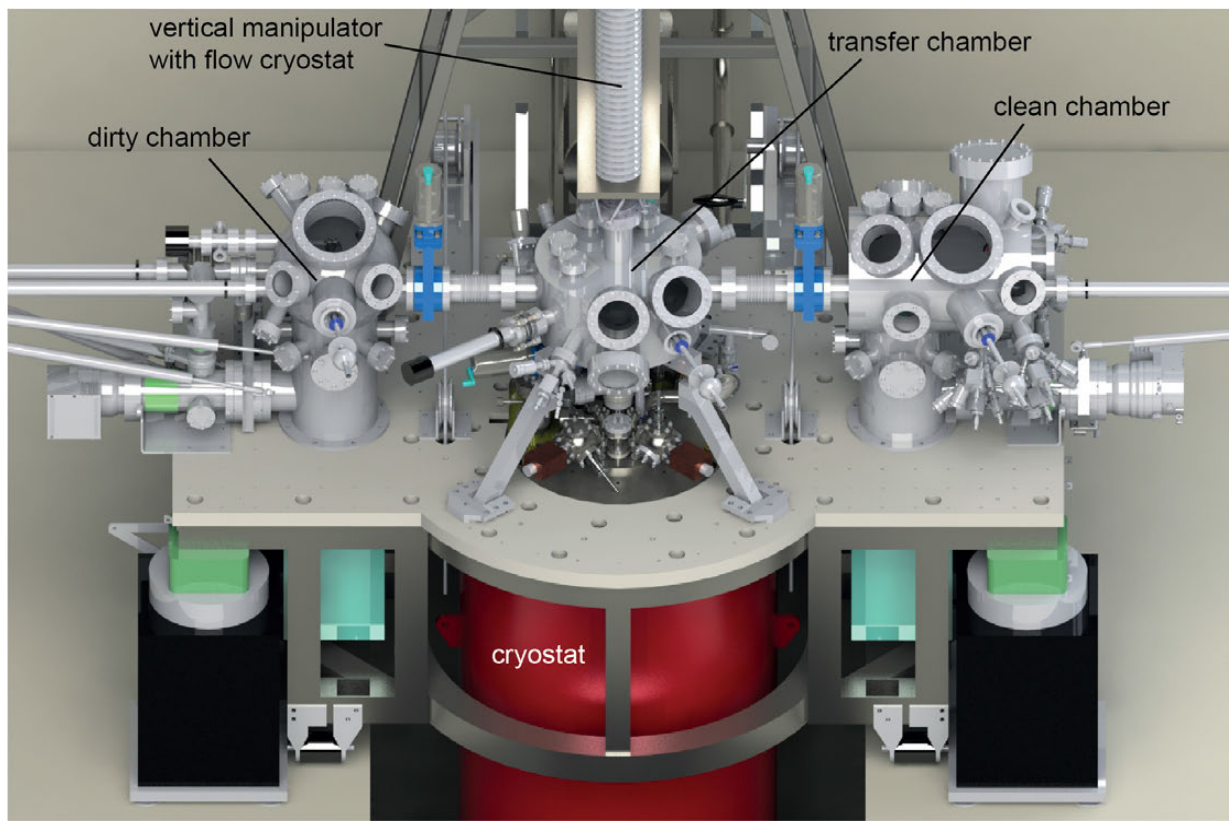

Figure 3. The UHV system comprises three interconnected chambers, separable by valves. The cryostat (red) is below the central vessel. The two satellite chambers are utilized for MBE and CVD preparation, as well as sputtering and annealing of samples.

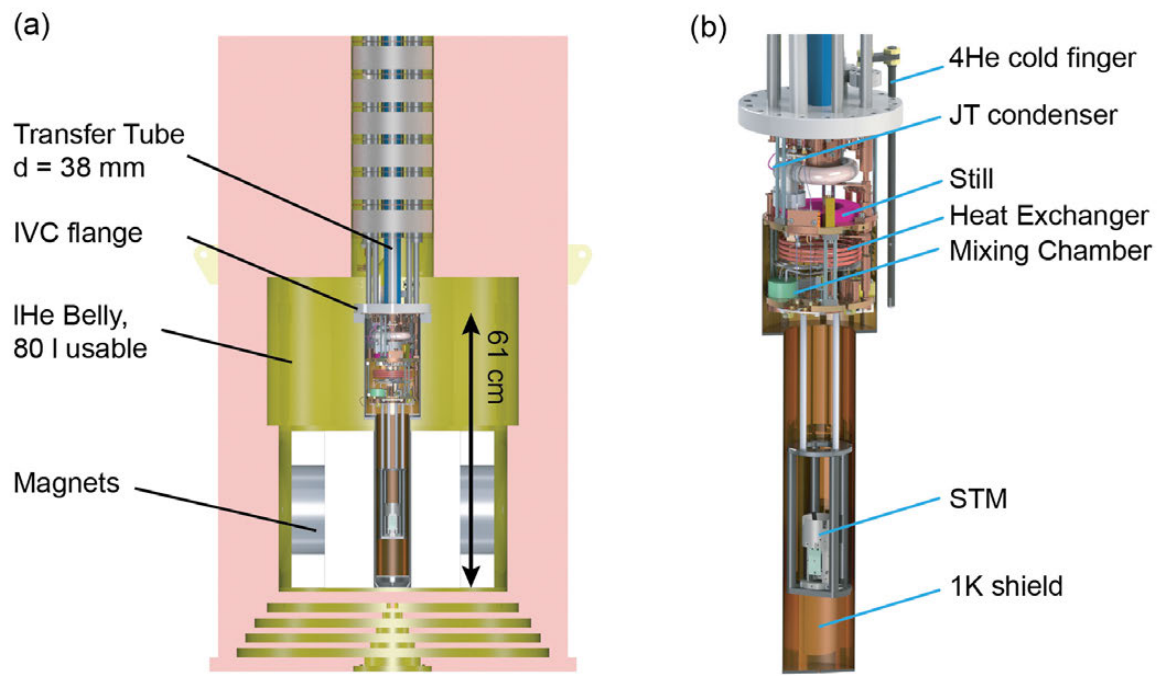

Figure 4. (a) Cut view of the top-loading cryostat. The liquid helium belly (yellow) contains the two separated magnets (solenoid for out-of-plane, split-coil for in-plane field) and surrounds the immersed inner vacuum can (IVC) containing the UHV-parts. All connections like electrical feedthroughs, magnet connectors and filling ports are placed at the top. (b) Detailed view on the dilution refrigerator parts placed inside UHV in the IVC, creating temperatures in the milli-Kelvin range. The STM is rigidly connected via gold-plated silver posts to the mixing chamber stage.

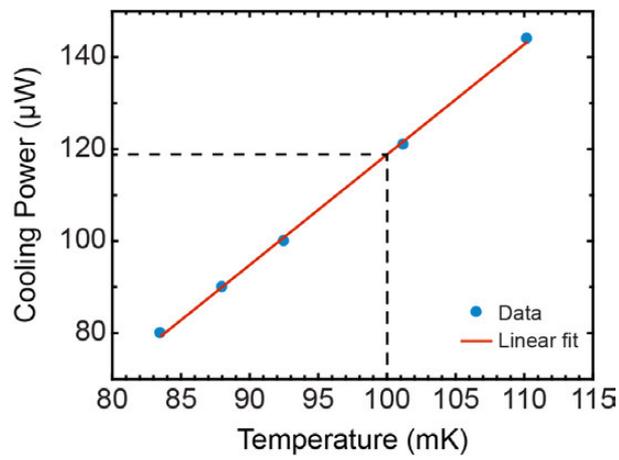

Figure 5. Cooling power of the JDR-50 dilution refridgerator as a function of the temperature measured at the mixing chamber without the microscope attached. At 100 mK the cooling power is 115 µW with a continuously running JT-compressor.

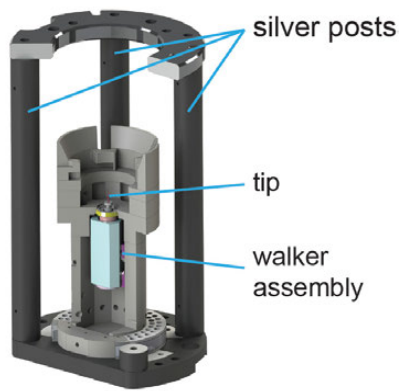

Figure 6. Cut view of the STM body and the silver posts that connect the microscope to the mixing chamber. The tip pointing upward can be moved by the piezo-driven stepper motor, utilizing the Pan design, towards the sample (latter not shown here).

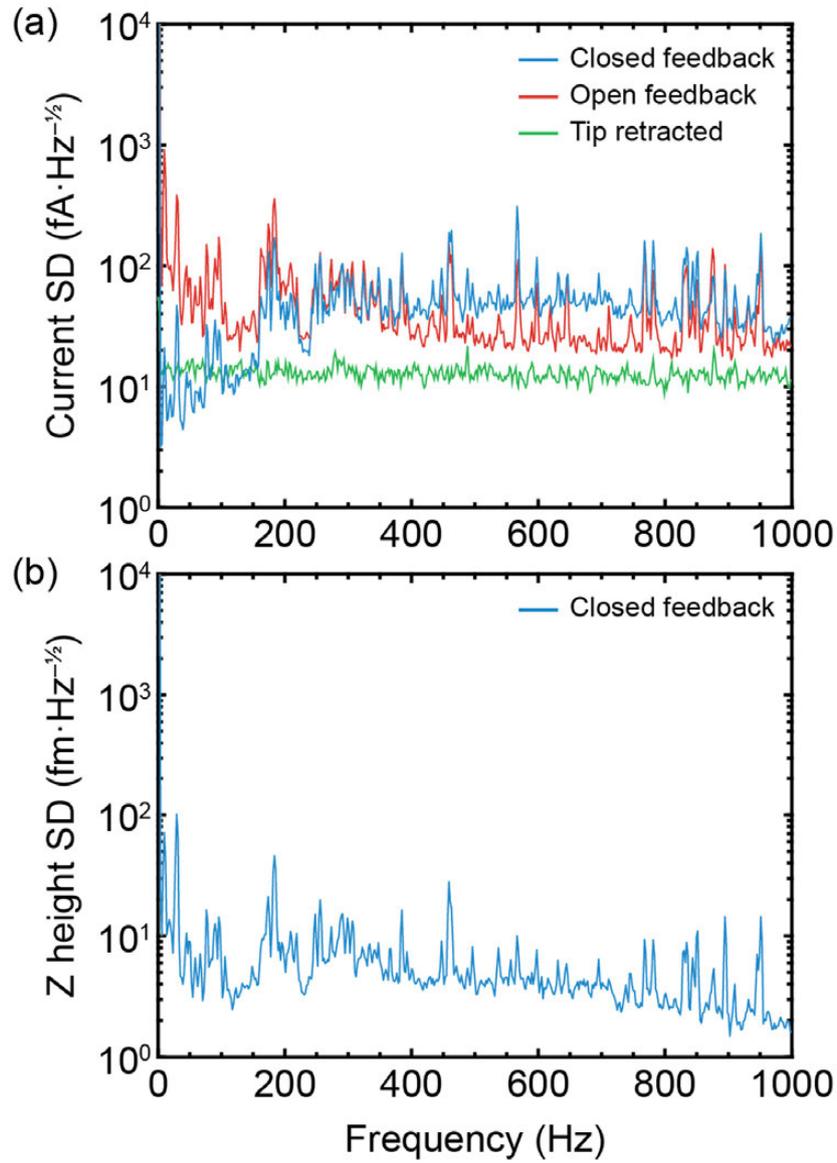

Figure 7. (a) Measured spectral density of the current signal for frequencies up to 1 kHz. Shown is the open (red) and closed (blue) feedback noise compared to the situation of a retracted tip (green), i.e. out of tunneling range ($I_t$ = 500 pA, $V_S$ = 10 mV, $T_{MC}$ = 43 mK). The data were taken with a PtIr tip on a thick Pb film grown on W(110).
(b) Measured spectral density for the z-signal under the same measuring conditions.

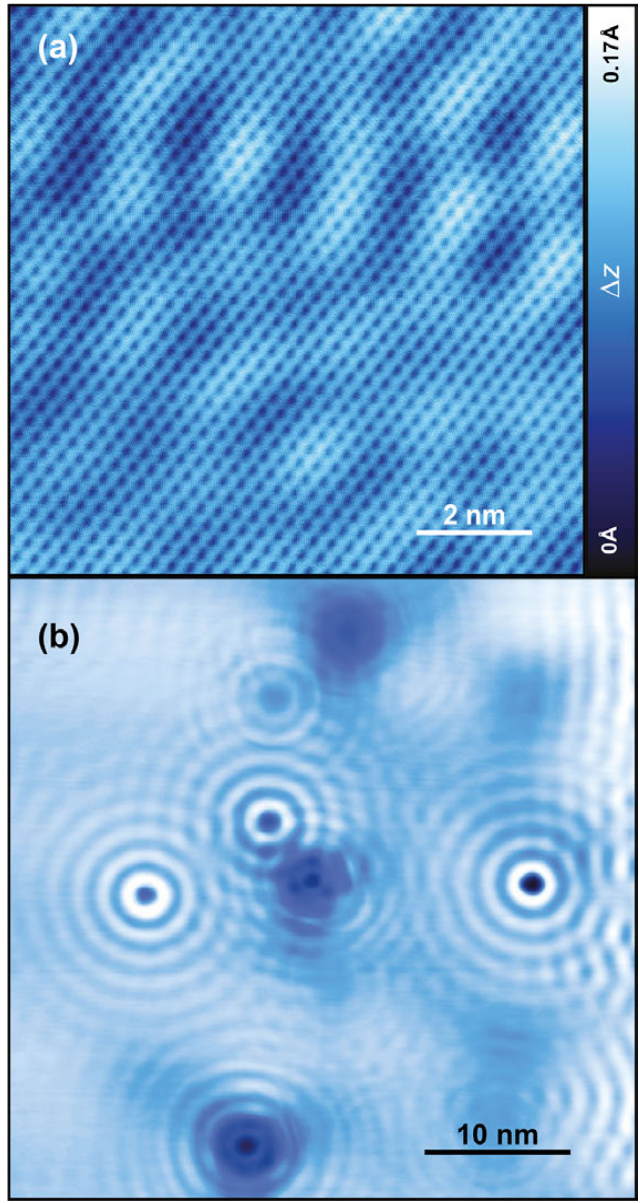

Figure 8. Qualitative proof of the instrument performance at base temperature $T_{MC}$ = 29 mK. (a) Atomic resolution of Cu(111) on a 10 nm x 10 nm area ($I_t$ = 10 nA, $V_S$ = 12 mV). (b) Measured quasi-particle interference (QPI) on the Cu(111) surface ($I_t$ = 200 pA, $V_S$ = 100 mV). The measurements were done with a PtIr tip.

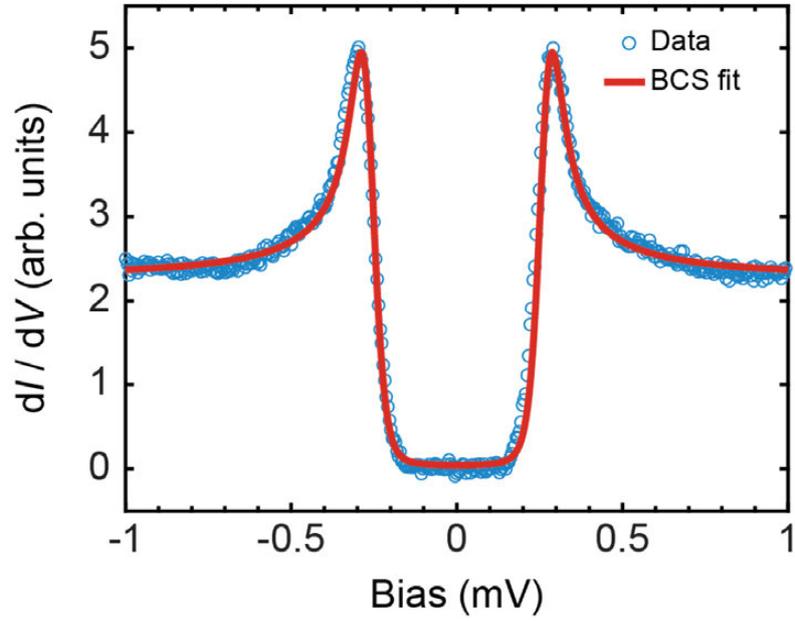

Figure 9. The measurement of the differential tunneling conductance on Re(0001) with a PtIr tip, using a lock-in amplifier with an AC modulation of $V_{mod}$ = 4 µV at $f_{mod}$ = 821 Hz, shows a superconducting gap ($T_{MC}$ = 47.5 mK). A BCS fit reveals an energy resolution (effective electron temperature) of $T_{eff}$ = 195 mK, assuming a gap $\Delta_{Re}$ = 0.265 mV and a lifetime broadening $\Gamma$ = 0.005 mV. The data shown is an average of 30 consecutive measurements.

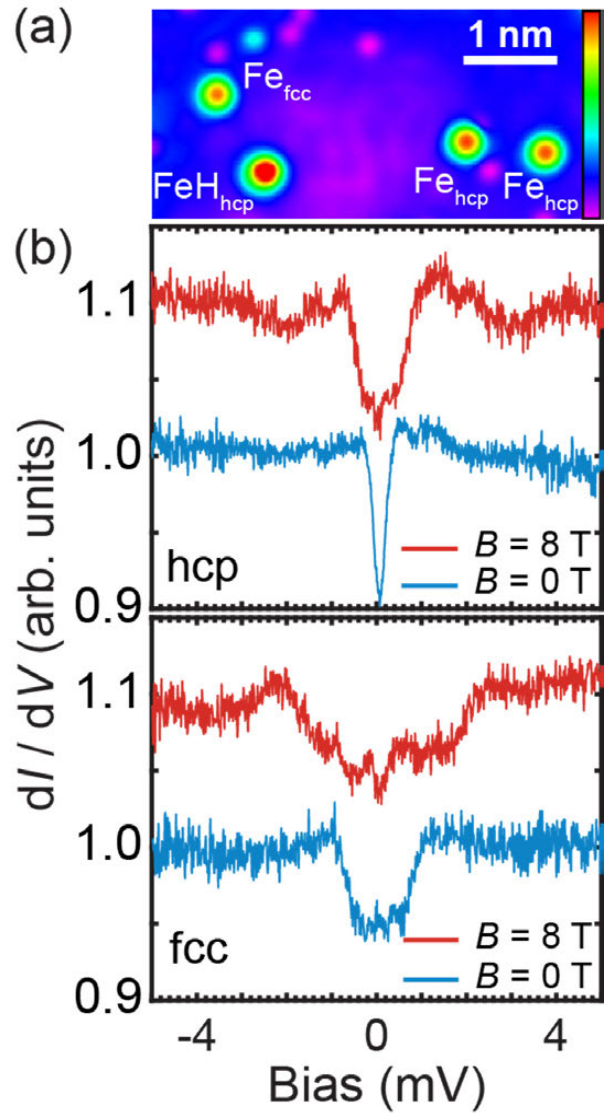

Figure 10. Individual clean Fe atoms and a hydrogenated Fe atom on a bare Pt(111) surface, after cold deposition ($T$ = 4.3 K), located on hcp/fcc hollow sites, as labeled ($I_t$ = 25 pA, $V_S$ = 10 mV, $T_{MC}$ = 37 mK). The color scale represents $\Delta z = 1.76$ Å. (b) Measured spin excitations of clean Fe adatoms adsorbed on an hcp hollow site (top) and fcc hollow site (bottom) on Pt(111) at the indicated out-of-plane magnetic field $B$. The spectra were normalized to the substrate and are offset for clarification. Stabilization values: $V_S$ = 5 mV, $I_t$ = 500 pA, $V_{mod}$ = 60 μV, $T_{MC}$ = 37 mK.

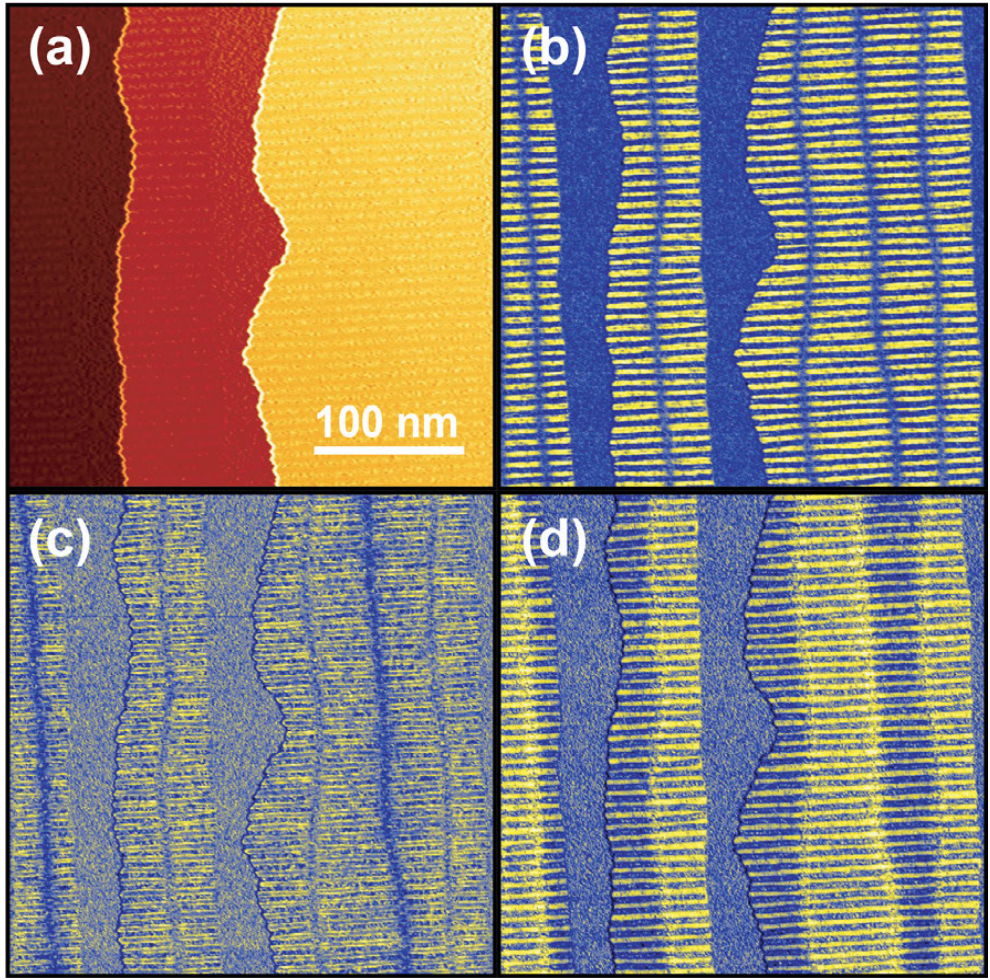

Figure 11. Spin-polarized STM of the double layer of Fe on W(110) imaged with a bulk Cr tip with a canted out-of-plane magnetization in constant-current mode. (a) Topography ($V_S$ = 50 mV, $I_t$ = 300 pA). The contrast variation corresponds to $\Delta z$ = 0.7 nm. (b)-(d) d$I$/d$V$ images of (b) the domain walls ($V_S$ = 50 mV, $I_t$ = 300 pA), (c) the in-plane contrast of the domain walls ($V_S$ = 15 mV, $I_t$ = 300 pA), and (d) the out-of-plane contrast of the magnetized domains ($V_S$ = -10 mV, $I_t$ = 300 pA). All images are the same size. ($T_{MC}$ = 33 mK, $f_{mod}$ = 4.09 kHz, $V_{mod}$ = 15 mV$_{rms}$).